\documentclass[onecolumn,a4paper]{IEEEtran}
\usepackage{amsmath}
\usepackage{amssymb}
\usepackage{amsfonts}
\usepackage{epsfig}
\usepackage{subfigure}
\usepackage{calc}
\usepackage{color}
\usepackage[all]{xy}
\usepackage{bm}
\usepackage{algorithmicx}
\usepackage{algpseudocode}
\usepackage{algorithm}
\usepackage{enumerate}
\usepackage[noadjust]{cite}
\usepackage{caption}
\usepackage{cancel}
\usepackage{soul}

\newtheorem{defn}{Definition}
\newtheorem{thm}{Theorem}[section]
\newtheorem{cor}[thm]{Corollary}
\newtheorem{prop}{Proposition}

\newtheorem{lem}[thm]{Lemma}
\newtheorem{conj}[thm]{Conjecture}
\newtheorem{constr}[thm]{Construction}
\newtheorem{note}{Remark}
\newtheorem{example}{Example}
\newcommand{\bit}{\begin{itemize}}
\newcommand{\eit}{\end{itemize}}
\newcommand{\bcor}{\begin{cor}}
\newcommand{\ecor}{\end{cor}}
\newcommand{\beq}{\begin{equation}}
\newcommand{\eeq}{\end{equation}}
\newcommand{\beqn}{\begin{equation*}}
\newcommand{\eeqn}{\end{equation*}}
\newcommand{\bea}{\begin{eqnarray}}
\newcommand{\eea}{\end{eqnarray}}
\newcommand{\bean}{\begin{eqnarray*}}
\newcommand{\eean}{\end{eqnarray*}}
\newcommand{\ben}{\begin{enumerate}}
\newcommand{\een}{\end{enumerate}}
\newcommand{\bdefn}{\begin{defn}}
\newcommand{\edefn}{\end{defn}}
\newcommand{\bnote}{\begin{note}}
\newcommand{\enote}{\end{note}}
\newcommand{\bprop}{\begin{prop}}
\newcommand{\eprop}{\end{prop}}
\newcommand{\blem}{\begin{lem}}
\newcommand{\elem}{\end{lem}}
\newcommand{\bthm}{\begin{thm}}
\newcommand{\ethm}{\end{thm}}
\newcommand{\bconj}{\begin{conj}}
\newcommand{\econj}{\end{conj}}
\newcommand{\bconstr}{\begin{constr}}
\newcommand{\econstr}{\end{constr}}
\newcommand{\bpf}{\begin{proof}}
\newcommand{\epf}{\end{proof}}

\begin{document}
\sloppy
\title{Codes With Hierarchical Locality}

\author{
	  \IEEEauthorblockN{Birenjith Sasidharan, Gaurav Kumar Agarwal, and P. Vijay Kumar} \\
	  \IEEEauthorblockA{Department of Electrical Communication Engineering, Indian Institute of Science, Bangalore.\\
	    Email: {biren, agarwal, vijay}@ece.iisc.ernet.in \thanks{This research is supported in part by the National Science Foundation under Grant No. 1422955 and in part by the joint UGC-ISF Research Grant No. 1676/14. Birenjith Sasidharan would like to acknowledge the support
of TCS Research Scholar Programme Fellowship.}} 
}
\maketitle

\begin{abstract}
In this paper, we study the notion of {\em codes with hierarchical locality} that is identified as another approach to local recovery from multiple erasures.  The well-known class of {\em codes with locality} is said to possess hierarchical locality with a single level. In a {\em code with two-level hierarchical locality}, every symbol is protected by an inner-most local code, and another middle-level code of larger dimension containing the local code. We first consider codes with two levels of hierarchical locality, derive an upper bound on the minimum distance, and provide optimal code constructions of low field-size under certain parameter sets. Subsequently, we generalize both the bound and the constructions to hierarchical locality of arbitrary levels. 
\end{abstract}
\begin{IEEEkeywords} Codes with locality, locally recoverable codes, hierarchical locality, multiple erasures, distributed storage.
\end{IEEEkeywords}

\section{Introduction\label{sec:intro}}
An important desirable attribute in a distributed storage system is the efficiency in carrying out repair of failed nodes. Among many others, two important metrics to characterize efficiency of node repair are \textit{repair bandwidth}, i.e., the amount of data download in the case of a node failure and \textit{repair degree}, i.e., the number of helper nodes accessed for node repair. While regenerating codes~\cite{DimGodWuWaiRam} aim to minimize the repair bandwidth, codes with locality~\cite{GopHuaSimYek} seek to minimize the repair degree. The focus of the present paper is on codes with locality. 

\subsection{Codes with Locality}
An $[n,k,d]$ linear code $\mathcal{C}$ can possibly require to access $k$ symbols to recover one lost symbol. The notion of locality of code symbols was introduced in~\cite{GopHuaSimYek}, with the aim of designing codes in such a way that the number of symbols accessed to repair a lost symbol is much smaller than the dimension $k$ of the code. The code $\mathcal{C}$ is said to have locality $r$ if the $i$-th code symbol $c_i, 1\leq i \leq n$ can be recovered by accessing $r << k$ other code symbols. In~\cite{GopHuaSimYek}, authors proved an upper bound on the minimum distance of codes with locality, and showed that an existing family of pyramid codes~\cite{HuaCheLi} can achieve the bound. In~\cite{KamPraLalKum}, authors extended the notion to $(r,\delta)$-locality, where each symbol can be recovered {\em locally} even in the presence of an additional $(\delta-2)$ erasures. In~\cite{GopHuaSimYek}, authors introduced categories of {\em information-symbol} and {\em all-symbol} locality. In the former, local recoverability is guaranteed for symbols from an information set, while in the latter, it is guaranteed for every symbol. Explicit constructions for codes with all-symbol locality are provided in~\cite{SilRawVis_allerton},~\cite{TamPapDim}, respectively based on rank-distance and Reed-Solomon (RS) codes. Improved bounds on the minimum distance of codes with all-symbol locality are provided in \cite{PraLalKum_isit,WanZha_int}, along with certain optimal constructions. Families of codes with all-symbol locality with small alphabet size (low field size) are constructed in~\cite{TamBar}.  Locally repairable codes over binary alphabet are constructed in~\cite{GopCal_isit}. A new approach of {\em local regeneration}, where in repair is both local and in addition bandwidth-efficient within the local group, achievable by making use of a vector alphabet is considered in~\cite{KamPraLalKum,SilRawKoyVis_isit,HuaBiePen}.

Recently, many approaches are proposed in literature~\cite{KamPraLalKum,PraLalKum_isit,TamBar,WanZha,PamHolOgg_isit} to address the problem of recovering from multiple erasures locally. The notion of $(r,\delta)$-locality introduced in~\cite{KamPraLalKum} is one such. In~\cite{WanZha}, an approach of protecting a single symbol by multiple support-disjoint local codes of the same length is considered. An upper bound on the minimum distance is derived, and existence of optimal codes is established under certain constraints. A similar approach is considered in \cite{TamBar} also. In \cite{TamBar}, authors allow {\em multiple recovering sets} of different sizes, and also provide constructions requiring field-size only in the order of block-length. Quite differently, authors of~\cite{PraLalKum_isit} consider codes allowing sequential recovery of two erasures, motivated by the fact that such a family of codes allow a larger minimum distance. An upper bound on the minimum distance and optimal constructions for restricted set of parameters are provided. 

\subsection{Our Contributions}
In the present paper, we study the notion of {\em hierarchical locality} that is identified as another approach to local recovery from multiple erasures. In consideration of practical distributed storage systems, Duminuco et al. in \cite{DumBie} had proposed the topology of {\em hierarchical codes} earlier. They compared hierarchical codes with RS codes in terms of repair-efficiency using real network-traces of KAD and PlanetLab networks. Their work was focused on collecting empirical data for performance improvements, rather than undertaking a theoretical study of such a topology. In the present paper, we study codes with hierarchical locality, first considering the case of two-level hierarchy. We derive an upper bound on the minimum distance and provide optimal code constructions under certain parameter-sets. This is further generalized to a setting of {\em $h$-hierarchy} in a straightforward manner. 

\section{Codes with Hierarchical Locality\label{sec:chl-2}}
\begin{figure}
	\centering
	\captionsetup{justification=centering}
	\includegraphics[height=0.5in]{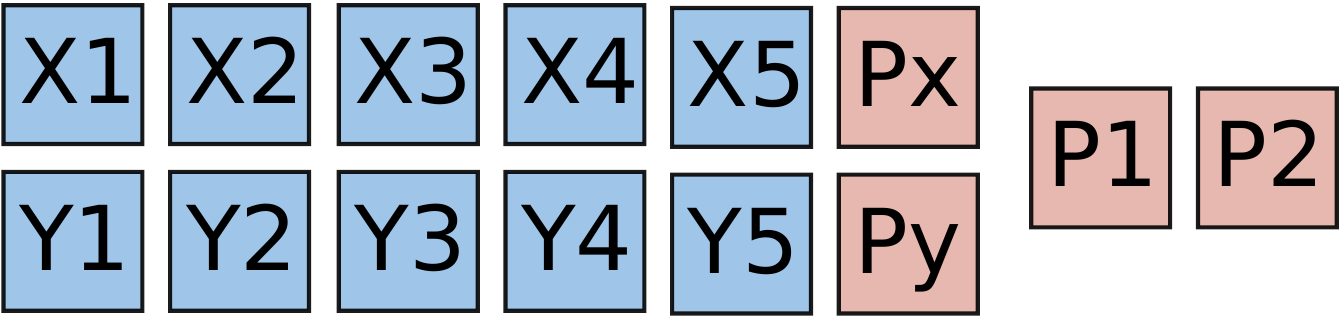} 
	\caption{Illustration of $[16,12,4]$-code used in Windows Azure.\label{fig:azure}} 
\end{figure}

The Windows Azure Storage solution employs a $[16,12,4]$-pyramid code with a locality parameter $r=6$. In the code, as illustrated in Fig.~\ref{fig:azure}, every code symbol except the global parities $P_1, P_2$ can be recovered accessing $r=6$ other code symbols.
While the code performs well in systems where single node-failure remains the dominant event, it requires to connect to $k=12$ symbols to recover a failed under certain erasure-patterns consisting of $2$ node-failures. We consider an example of $[24,14,6]$-code from the family of {\em codes with hierarchical locality} in an attempt to reduce such an overhead. 
\begin{figure}
	\centering
	\captionsetup{justification=centering}
	\includegraphics[width=3in]{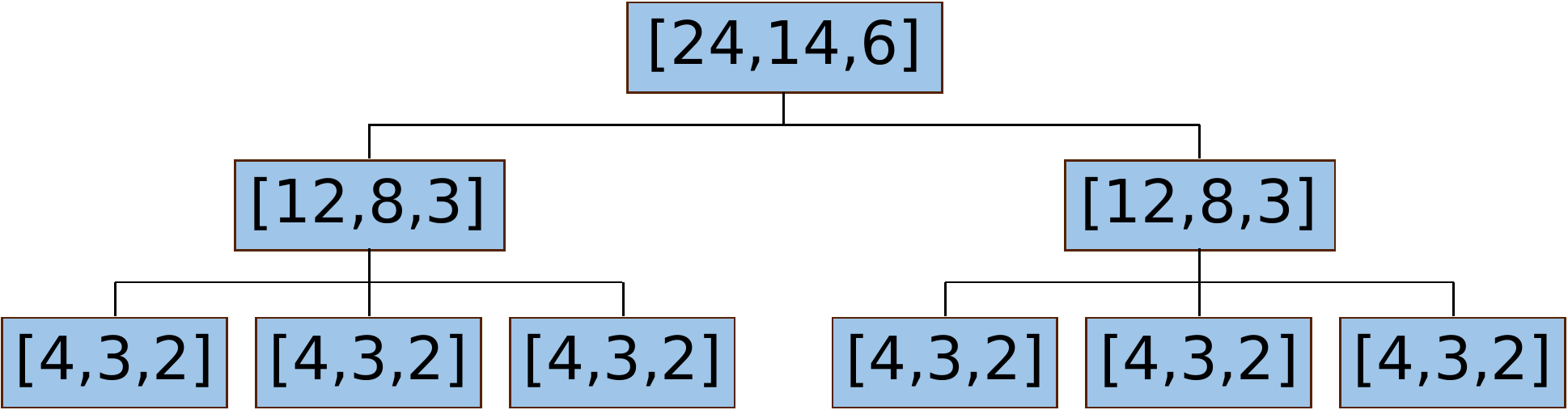} 
	\caption{Illustration of $[24,14,6]$-code having $2$-level hierarchical locality.\label{fig:topology}} 
\end{figure}
The structure of the code is depicted in Fig.~\ref{fig:topology} as a tree in which each node represents a constituent code. The code contains two support-disjoint $[n_1=12,r_1=8,d_2=3]$ codes, each of them in turn comprised of three support-disjoint $[n_2=4,r_2=3,d_1=2]$ codes. Making use of $[4,3,2]$-code, all single-erasures can be repaired accessing $r_2=3$ symbols, which is half the number of symbols required in the Windows Azure code in a similar situation. We can recover a lost symbol connecting to $r_1=8$ symbols in the case of erasure-pattern involving $2$ erasures. This is in contrast to the Windows Azure code where we had to download the entire message of $12$ symbols. While the Windows Azure code offers a storage overhead of $1.3$x with a minimum distance of $4$, our code has a larger overhead of $1.7$x with a better minimum distance $d=6$. The example of $[24,14,6]$-code can indeed be constructed, and it will be shown that the minimum distance is optimal among the class of codes.

\subsection{Preliminaries \label{sec:prelim}}
\bdefn \label{def:cl}\cite{KamPraLalKum} An $[n,k,d]$ linear code $\mathcal{C}$ is a code with $(r,\delta)$-locality if for every symbol $c_i, 1 \leq i \leq n$, there exists a punctured code $C_i$ such that $c_i \in \textsl{Supp}(C_i)$ and the following conditions hold: 1) $\textsl{dim}(C_i) \leq r$, 2) $d_{\text{min}}(C_i) \geq \delta$.
\edefn

Codes with locality were first defined in \cite{GopHuaSimYek} for the case of $\delta = 2$, and the class was generalized for arbitrary $\delta$ in \cite{KamPraLalKum}. In the definition given in \cite{KamPraLalKum}, the authors imposed constraints on the length and the $d_{\text{min}}$ of $C_i$. We replace the constraint on length with a constraint on $\textsl{dim}(C_i)$, and it may be noted that it does not introduce any loss in generality. The code $C_i$ associated with the $i$-the symbol is referred to as its {\em local code}. If it is sufficient to have local codes only for symbols belonging to some fixed information set $I$, such codes are referred to as {\em codes with information-symbol $(r, \delta)$-locality}. The general class in Def.~\ref{def:cl} is also referred to as {\em codes with all-symbol $(r,\delta)$-locality}, in order to differentiate them from the former. In this paper, unless otherwise mentioned, we consider codes with all-symbol locality.
\bdefn \label{def:chl-2} An $[n,k,d]$ linear code $\mathcal{C}$ is a {\em code with hierarchical locality} having locality parameters $[(r_1,\delta_1), (r_2,\delta_2)]$ if for every symbol $c_i, 1 \leq i \leq n$, there exists a punctured code $C_i$ such that $c_i \in \textsl{Supp}(C_i)$ and the following conditions hold: 1) $\textsl{dim}(C_i) \leq r_1$, 2) $d_{\text{min}}(C_i) \geq \delta_1$, 3) $C_i$ is a code with $(r_2,\delta_2)$-locality.
\edefn

The punctured code $C_i$ associated with $c_i$ is referred to as its {\em middle code}. Since the middle code is a code with locality, each of its symbols will in turn be associated with a local code. 

\subsection{An Upper Bound On the Minimum Distance \label{sec:bound}}
\bthm \label{thm:bound} Let $\mathcal{C}$ be an $[n,k,d]$-linear code with hierarchical locality having locality parameters $[(r_2,\delta_2), (r_1,\delta_1)]$. Then
\bea \label{eq:bound}
d \leq n-k+1 - \left( \left\lceil \frac{k}{r_2} \right\rceil -1 \right)(\delta_2 -1) - \left(\left\lceil \frac{k}{r_1} \right\rceil -1 \right)(\delta_1 - \delta_2).
\eea
\ethm
\bpf We extend the techniques introduced in \cite{GopHuaSimYek} in proving the theorem. A punctured code $\mathcal{C}_s$  of $\mathcal{C}$ having dimension $k-1$, is identified first. Then we will use the fact that

\bea \label{eq:fact}
d & \leq & n - |\textsl{Supp}(\mathcal{C}_s)|.
\eea

The Algorithm~\ref{alg:bound} (see flow chart in App.~\ref{app:flowcharts}) in is used to find $\mathcal{C}_s$ with a large support. In each iteration indexed by $j$, the algorithm identifies a middle code from $\mathcal{C}$, that accumulates additional rank. Then it picks up local codes from within the middle code that accumulate additional rank. Clearly, the algorithm terminates as the total rank is bounded by $k$. Let $i_{\text{end}}$ and $j_{\text{end}}$ respectively denote the final values of the variables $i$ and $j$ before the algorithm terminates. Let $a_i$ denote the incremental rank and $s_i$ denote the incremental support while adding a local code $L_i$. Then we have $s_i \geq  a_i + (\delta_2 -1), \ \ 1 \leq i \leq i_{\text{end}}$, since we have $a_i > 0$ in every iteration. The set $S_i$ denotes the support of $L_i$, and $V_i$ denotes the space $\textsl{Column-Space}(G|_{S_i})$, where $G$ is the generator matrix of the code. If no more local codes can be added from the middle code $M_j$, then the support of the last local code added from $M_j$ is removed and an additional support $T_j$ of $M_j$ is added to $\Psi$. Let $i(j)$ denote the index of the last local code added from $M_j$. Since the middle code has a minimum distance of $\delta_1$, and every rank accumulating local code brings at least one new information symbol, it follows that

\bean
t_j := |T_j| & \geq & a_{i(j)} + (\delta_1 -1), \\
& = & a_{i(j)} + (\delta_2 -1) +  (\delta_1 -\delta_2) \ \ 1 \leq j \leq j_{\text{end}} .
\eean

\begin{algorithm}
	\caption{For the proof of Thm.~\ref{thm:bound} \label{alg:bound}}
	\begin{algorithmic}[1]
		\State Let $j=0, i=0, W = \phi, \Psi= \phi$.
		\While {($\exists$ a middle code $M_j \in \mathcal{C}$ such that $\textsl{rank}(G|_{\Psi \cup M_j}) > \textsl{rank}(G|_{\Psi})$) }
			\While {($\exists$ a local code $L_i \in M_j$ such that $V_i \subsetneq W$)}
				\State $W = W+ V_i$
				\State $\Psi = \Psi \cup S_i$
				\State $i= i + 1$
			\EndWhile
			\State $\Psi = (\Psi \setminus S_{i-1}) \cup T_j$
			\State $j= j + 1$
		\EndWhile
	\end{algorithmic}
\end{algorithm} 
The rank accumulates to $k$ after adding the last local code $L_{i_{\text{end}}}$. We would also have visited $j_{\text{end}}$ middle codes by then. Hence,
\bea
\label{i_j_last}
i_{\text{end}}  \geq  \left\lceil \frac{k}{r_2} \right\rceil , \ \ j_{\text{end}} \geq  \left\lceil \frac{k}{r_1} \right\rceil .
\eea
After adding $L_{i_{\text{end}}-1}$ local codes, we would have accumulated rank that is less than or equal to $(k-1)$. Hence we can always pick $s_e := (k-1) - \sum_{i=1}^{i_{\text{end}}-1} a_i$ columns from $L_{i_{\text{end}}}$ so that the total rank accumulated becomes $(k-1)$. Note that $s_e \geq 0$. The resultant punctured code is identified as $\mathcal{C}_s$. Let $E = \{ i(j) \mid 1 \leq j \leq j_{\text{end}}\}$. Then

\beq \label{eq:total_support} 
|\textsl{Supp}(\mathcal{C}_s)| \ \geq \ \sum_{\substack{i \notin E, i=1}}^{i_{\text{end}}-1 } s_i + s_e + \sum_{j=1}^{j_{\text{end}}-1} t_j .
\eeq
In \eqref{eq:total_support}, the last term $\sum_{j=1}^{j_{\text{end}}-1} t_j$ includes a sum of only $j_{\text{end}}-1$ terms because we could have possibly accumulated a rank of $(k-1)$ after adding $L_{i_{\text{end}}-1}$, i.e., $s_e = 0$. Thus we have,
\bean 
|\textsl{Supp}(\mathcal{C}_s)| & \geq & \sum_{\substack{i \notin E,  i=1}}^{i_{\text{end}}-1} s_i + (k-1) - \sum_{i=1}^{i_{\text{end}}-1} a_i + \sum_{j=1}^{j_{\text{end}}-1} t_j \\
& \geq & \sum_{\substack{i \notin E,  i=1}}^{i_{\text{end}}-1} (a_i + (\delta_2 -1)) + (k-1) - \sum_{i=1}^{i_{\text{end}}-1} a_i \\
& & + \sum_{j=1}^{j_{\text{end}}-1}  (a_{i(j)} + (\delta_2 -1) +  (\delta_1 -\delta_2)) \\
& = & \sum_{i=1}^{i_{\text{end}}-1} (\delta_2 -1) + (k-1) + \sum_{j=1}^{j_{\text{end}}-1}  (\delta_1-\delta_2) \\ 
\eean
Substituting values of $i_{\text{end}}$ and $j_{\text{end}}$ from \eqref{i_j_last} and using \eqref{eq:fact}, we obtain the bound.
\epf

It may be noted that the theorem holds good even for codes with information-symbol hierarchical locality. 


\subsection{Code Constructions For Information-Symbol Locality\label{sec:pyramid_constrn}}

A straightforward extension of pyramid codes~\cite{HuaCheLi} is possible to construct optimal codes with information-symbol hierarchical locality. They achieve the bound in \eqref{alg:bound} if $r_2 \mid r_1 \mid k$. In this section, we illustrate the construction with an example, assuming $\delta_2=2$. The two-level hierarchical code described  here extends naturally to multiple-level hierarchy and yields optimal codes if $r_h \mid r_{h-1} \mid \cdots \mid r_1 \mid k$. 

The construction is built on a systematic MDS code with parameters $[k+d-1,k,d]$ with a generator matrix $G_{\text{mds}}$. Let  
\bean
G_{\text{mds}} & = & [ I_{k \times k} \mid Q_{k \times (d-1)}] \\
k & = & \alpha r_1 + \beta r_2 + \gamma, \ \ 0 \leq \beta r_2 +\gamma < r_1 \\
r_1 & = & \mu r_2 + \nu, \ \ 0 \leq \nu < r_2.
\eean
Partition $Q$ as
\bean
Q & = & \left[ \begin{array}{c|c} \begin{array}{c} Q_1 \\ Q_2 \\ \vdots \\ Q_{\alpha + 1} \end{array} &  Q^{\prime} \end{array} \right],
\eean
where $Q_i, 1 \leq i \leq \alpha$ is of size $r_1 \times (\delta_1-1)$, $Q_{\alpha +1}$ is of size $(\beta r_2 + \gamma) \times (\delta_1-1)$  and $Q^{\prime}$ is of size $(k \times (d-\delta_1))$. Further partition $Q_i, 1 \leq i \leq \alpha$ as
\bean
Q_i & = & \left[ \begin{array}{c|c} \begin{array}{c} R_{i1} \\ R_{i2} \\ \vdots \\ R_{i,\mu+1} \end{array} & R_i^{\prime} \end{array} \right],
\eean
where $R_{ij}, 1 \leq j \leq \mu$ is of size $r_2 \times 1$, $R_{i,\mu +1}$ is of size $\nu \times 1$  and $R_i^{\prime}$ is of size $(r_1 \times (\delta_1-2))$. At the same time $Q_{\alpha +1}$ is partitioned as
\bean
Q_{\alpha +1} & = & \left[ \begin{array}{c|c} \begin{array}{c} R_{\alpha+1,1} \\ R_{\alpha+1,2} \\ \vdots \\ R_{\alpha+1,\beta + 1} \end{array} & R_{\alpha +1}^{\prime} \end{array} \right],
\eean
where $R_{\alpha +1,j}, 1 \leq j \leq \beta$ is of size $r_2 \times 1$, $R_{\alpha +1,\beta +1}$ is of size $\gamma \times 1$  and $R_{\alpha+1}^{'}$ is of size $((\beta r_2 + \gamma) \times (\delta_1-2))$. Next, we can construct matrices
\bean
\hat{Q}_i & = & \left[ \begin{array}{c|c} \begin{array}{cccc} R_{i1} & & & \\ 
		& R_{i2} & \\
		&& \ddots & \\
		&&& R_{i,\mu+1} \end{array} & R_i^{\prime} \end{array} \right] \\
\hat{Q}_{\alpha +1} & = & \left[ \begin{array}{c|c} \begin{array}{cccc} R_{\alpha+1,1} & && \\ 
		& R_{\alpha+1,2} & & \\
		& & \ddots & \\
		& & & R_{\alpha+1,\beta + 1} \end{array} & R_{\alpha +1}^{'} \\
\end{array} \right] \\
\hat{Q} & = & \left[ \begin{array}{c|c} \begin{array}{cccc} \hat{Q}_1 & & & \\
		& \hat{Q}_2 & & \\ 
		& & \ddots & \\ 
		& & & \hat{Q}_{\alpha + 1} \end{array} &  Q^{\prime} \end{array} \right]
\eean
Let $J$ be defined as 
\bean
J & = & \left[ \begin{array}{cccc} I_{r_2} & & &  \\
	&  \ddots & &  \\
	& & I_{r_2} &  \\
	& & & I_{\nu} \end{array} \right].
\eean
where $I_{r_2}$ is repeated $\mu$ times. Finally we construct the generator matrix $G$ of the pyramid code as,
\bean
G & = & \left[ \begin{array}{c|c} \begin{array}{ccccccc} J & & & & & &  \\
		&  \ddots & & & & &  \\
		& & J & & & &  \\
		& & & I_{r_2} & & & \\
		& & & & \ddots   & & \\
		& & & & & I_{r_2}  & \\
		& & & & & & I_{\gamma} \end{array} & \hat{Q} \end{array} \right],
\eean
with $J$ being repeated $\alpha$ times, and $I_{r_2}$ repeated $\beta$ times. The resultant code has length
\bean
n & = & k+d-1 + \left( \left\lceil \frac{k}{r_1} \right\rceil \left\lceil \frac{r_1}{r_2} \right\rceil - 1 \right) + \left(\left\lceil \frac{k}{r_1} \right\rceil - 1 \right)(\delta_1 -2).
\eean
Clearly, minimum distance of the code corresponding to $G$ is greater than or equal to that of $G_{\text{mds}}$, which is $d$. Also, $G$ satisfies the property of information-symbol hierarchical locality by construction. Using the bound in \eqref{eq:bound}, the code is optimal if
\bean
\left\lceil \frac{k}{r_1} \right\rceil \left\lceil \frac{r_1}{r_2} \right\rceil & = & \left\lceil \frac{k}{r_2} \right\rceil.
\eean

\subsection{Code Constructions For All-Symbol Locality\label{sec:constrn}}

We assume a divisibility condition $n_2 \mid n_1 \mid n$. The construction is described in three parts. The first part involves identification of a suitable finite field $\mathbb{F}_{p^m}$, a partition of $\mathbb{F}_{p^m}^*$ and a set of polynomials in $\mathbb{F}_{p^m}[X]$ that satisfy certain conditions. We require that every polynomial evaluates to a constant within one subset in the partition, and evaluates to zero in all the remaining subsets. In the second part, we construct a code polynomial $c(X)$ from the message symbols with the aid of the suitably chosen polynomials. The code polynomial $c(X)$ is formed in such a way that the locality constraints are satisfied. This part also involves precoding of message symbols in such a way that the dimension of the middle codes and the global code are kept to the desired values. Finally, the third part involves evaluation of the code polynomial at $n$ points of $\mathbb{F}_{p^m}^*$, that are chosen in the first part. 

\subsubsection{Identification of $\mathbb{F}_{p^m}$, a partition of $\mathbb{F}_{p^m}^*$ and a set of polynomials}
Let the finite field $\mathbb{F}_{p^m}$ be such that $n_1 \mid p^m - 1$, and $n < p^m$. Existence of such a pair $(p,m)$ is shown in App.~\ref{app:num_theory_prime}. We define the integers 
\bean
n_0 = p^m - 1, \mu_0 \ = \ 1,\mu_1 \ = \ \frac{n_0}{n_1}, \mu_2 \ = \ \frac{n_1}{n_2}.
\eean
Let $\alpha$ denote the primitive element of $\mathbb{F}_{p^m}$, and hence $\mathbb{F}^*_{p^m} \ = \ \{1, \alpha, \alpha^2, \ldots, \alpha^{p^m-2} \}$. Set $\beta_0 = \alpha$ and $\beta_1, \beta_2$ be elements of order $n_1$ and $n_2$ respectively. Then we have the following subgroups:
\bean
H_0 & = & \mathbb{F}^*_{p^m} \\
H_1 & = & \{1, \beta_1, \beta_1^2, \ldots, \beta_1^{n_1-1} \} \\
H_2 & = & \{1, \beta_2, \beta_2^2, \ldots, \beta_2^{n_2-1} .
\eean
We can further write 
\bean
H_0 & = & H_1 \uplus \beta_0 H_1 \uplus \cdots \uplus \beta_0^{\mu_1-1} H_1 \\
H_1 & = & H_2 \uplus \beta_1 H_1 \uplus \cdots \uplus \beta_1^{\mu_2-1} H_2 .
\eean

\begin{figure}[h]
\centering
\captionsetup{justification=centering}
\includegraphics[width=4in]{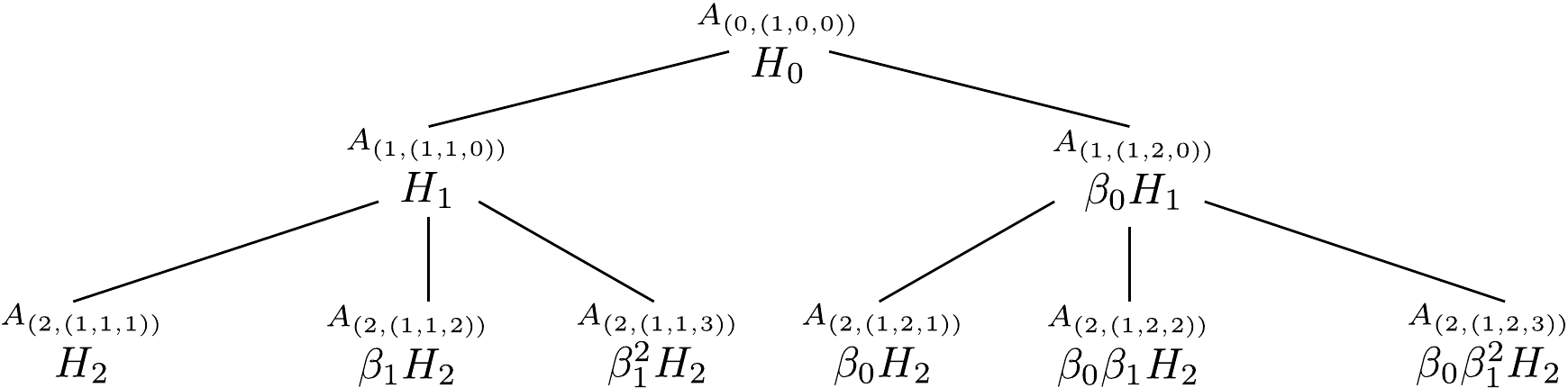} 
\caption{Illustration of the coset-tree. \label{fig:partition}} 
\end{figure}

Having set up a subgroup chain, we proceed to define a family of subsets of $H_0$. These subsets are indexed by a tuple $(i,\underline{t})$ with $i \in \{0,1,2\}$. For a given value of $i$, $\underline{t}$ takes values from the set $T_i = \{ (t_0, t_1, t_2) \mid 1 \leq t_j \leq \mu_j \text{ for } j \leq i; t_j = 0 \text{ for } j > i \}$. For a given tuple $(i,\underline{t})$, let us define a coset $A_{(i,\underline{t})}$ of the subgroup $H_i$ as follows:
\bean
\gamma_{(i,\underline{t})} =  \prod_{j=0}^{i-1} \beta_j^{t_{j+1} -1} , \ \ \ A_{(i,\underline{t})} =  \gamma_{(i,\underline{t})} H_i.
\eean
The set of possible indices has a tree-structure with each index $(i,\underline{t})$ associated with a unique vertex of the tree. A vertex $(i,\underline{t})$ belongs to the $i$-th level of the tree, and the $3$-tuple $\underline{t}$ describes the unique path from the vertex to the root of the tree. The parent of a vertex $(i,\underline{t})$ is denoted by $\pi(i,\underline{t})$, and the set of its siblings, i.e, other vertices having the same parent, is denoted by $\psi(i,\underline{t})$. The tree structure of $\{ A_{(i,\underline{t})} \mid i \in \{0,1,2\}, \underline{t} \in T_i\}$ is depicted in Fig.~\ref{fig:partition}. Since each vertex at $i$-th level is associated with a coset of $H_i$, we refer to this tree as the {\em coset-tree}.

Next, we define the polynomials $p_{(i,\underline{t})}(X), q_{(i,\underline{t})}(X) \in \mathbb{F}_{p^m}[X]$ as, 
\bean
p_{(i,\underline{t})}(X)  = \prod_{\theta \in A_{(i,\underline{t})}} (X - \theta), \ \ q_{(i,\underline{t})}(X) = \prod_{\underline{s} \in \psi(i,\underline{t})} p_{(i,\underline{s})} (X) .
\eean
The polynomial $p_{(i,\underline{t})}(X)$ is the annihilator of $A_{(i,\underline{t})}$. Furthermore, the polynomials in $\{ q_{(i,\underline{t})}(X) \mid 1 \leq t_i \leq \mu_i, t_j \text{ is fixed for } j  \neq i \}$ are relatively prime collectively. Thus for $i = 1, 2$ there exists $\{ a_{(i,\underline{t})}(X) \mid 1 \leq t_i \leq \mu_i, t_j \text{ is fixed for } j  \neq i \}$ such that
\bea 
\sum_{\substack{t_i = 1 \\ t_j \text{fixed for} j \neq i}}^{\mu_i}a_{(i,\underline{t})}(X) q_{(i,\underline{t})}(X)  =  1 \ \text{mod} \ (X^{n_{i-1}} - \gamma_{(i-1,\underline{s})}^{n_{i-1}} ), \label{eq:gcd}
\eea
where $\underline{s} = \pi(i,\underline{t})$. Next, we define $E_{(i,\underline{t})}(X) = a_{(i,\underline{t})}(X) q_{(i,\underline{t})}(X)$, and determine a valid candidate for $E_{(i,\underline{t})}(X)$ in the next Lemma such that~\eqref{eq:gcd} holds. Subsequently in Lem.~\ref{lem:poly_props}, we will list down certain useful properties of these polynomials. The proof is relegated to Appendix.
\blem \label{lem:poly}  
\beq E_{(i,\underline{t})}(X) = \prod_{\underline{s} \in \psi(i,\underline{t})} \left[ \frac{X^{n_i} - (\gamma_{(i,\underline{s})})^{n_i}}{(\gamma_{(i,\underline{t})})^{n_i} - (\gamma_{(i,\underline{s})})^{n_i}} \right]. \eeq
\elem
\blem \label{lem:poly_props} Let $i \in \{1, 2\}$, $\underline{t}, \underline{s}  \in T_i$ and $\underline{t} \in \psi(i,\underline{s})$. Let $\underline{\tau} = (\tau_0, \tau_1,\tau_2) = \pi(i,\underline{t})$. Then
\bea
\label{eq:prop1} E_{(i,\underline{t})}(X) & = & g(X^{n_i}) \text{ for some polynomial } \\
\nonumber & &   g(.), \ \textsl{deg}(g) = \mu_i - 1  \\
\label{eq:prop2} E_{(i,\underline{t})} (\theta) & = & \left\{ \begin{array}{cl} 1 & \theta \in A_{(i,\underline{t})}(X) \\
	0 & \theta \in \gamma_{(i-1,\underline{\tau})} H_{i-1} \setminus A_{(i,\underline{t})} \end{array} \right. \\
\label{eq:prop3} E_{(i,\underline{s})}(X) E_{(i,\underline{t})}(X)  & \equiv & 0 \ \text{mod} \ (X^{n_{i-1}} - \gamma_{(i-1,\underline{\tau})}^{n_{i-1}}) \\
\label{eq:prop4} E^2_{(i,\underline{t})}(X) = E_{(i,\underline{t})}(X) &\text{mod}& (X^{n_{i-1}} - \gamma_{(i-1,\underline{\tau})}^{n_{i-1}}).
\eea
\elem
\bpf The property~\eqref{eq:prop1} is clear from the definition of $E_{(i,\underline{t})}(X)$. The properties~\eqref{eq:prop2}, \eqref{eq:prop3} are clear from the proof of Lemma~\ref{lem:poly}. Hence \eqref{eq:prop4} follows by~\eqref{eq:gcd}.
\epf
\subsubsection{Construction of $c(X)$} We start with associating message polynomials of degree $(r_2-1)$ with certain leaves of the coset-tree. The total number of leaves of the coset-tree equals $\mu_1\mu_2$. However, we will only consider a suitable subtree of the coset-tree such that the number of leaves equals $\bar{\mu}_1\mu_2$ where $\bar{\mu}_1 = \frac{n}{n_1}$. The required subtree is obtained by removing the last $(\mu_1 - \bar{\mu}_1)$ branches emanating from the root of the tree. Every leaf that is retained in the subtree has an index $(2,\underline{t})$ where $\underline{t}$ belongs to the set 
\bean
T^\prime_2 = \{ \underline{t} \in T_2 \mid 1 \leq t_1 \leq \bar{\mu}_1 \}.
\eean
This subtree is referred to as the {\em relevant coset-tree}. A vertex from the $i$-th level, $i > 0$ of the relevant coset-tree will have an index $(i,\underline{t})$ where $\underline{t} \in T^\prime_i = \{ \underline{t} \in T_i \mid 1 \leq t_1 \leq \bar{\mu}_1 \}$. 

Consider a set ${\cal U} = \{ u_{\underline{t}}(X) = u_{\underline{t},0} + u_{\underline{t},1} X + \ldots + u_{\underline{t},r_2-1} X^{r_2-1}  \mid \underline{t} \in T^\prime_2 \}$ of message polynomials of size $\bar{\mu}_1\mu_2$. The code polynomial $c(X)$ is built from ${\cal U}$ in an iterative manner. In every iteration, we take as input a set of polynomials corresponding to vertices of the $i$-th level of the relevant coset-tree, and output another set of polynomials corresponding to vertices of the $(i-1)$-th level. As noted earlier, each leaf of the relevant coset-tree is uniquely mapped to a polynomial in ${\cal U}$. In the end, we will identify a polynomial $c_{(0,(1,0,0))} (X)$ associated with the root of the relevant coset-tree. The code polynomial $c(X) = c_{(0,(1,0,0))} (X)$. It may be noted that the polynomials in ${\cal U}$ is made up of $\bar{\mu}_1\mu_2r_2$ message symbols in total. However, the desired dimension $k$ can be less than $\bar{\mu}_1\mu_2r_2$. Hence in every iteration, a precoding of message symbols is carried out causing a reduction in the number of independent message symbols. The dimension would be reduced to the desired value $k$ at the end of the final iteration. 

Let us now start the iteration by setting $ c_{(2,\underline{t})}(X) = u_{\underline{t}}(X) \ \ \forall  \underline{t} \in T^\prime_2$.
Evaluations of $c_{(2,\underline{t})}(X)$ at $n_2$ points in $A_{(2,\underline{t})}(X)$ give rise to an $[n_2,r_2]$-codeword. Recognizing this correspondence, we refer to $c_{(2,\underline{t})}(X), \underline{t} \in T^\prime_2$ as a {\em second level code polynomial}. In the next iteration, for every $\underline{t} \in T^\prime_1$
\bean
d_{(1,\underline{t})}(X) = \sum_{\underline{s}:\pi(2,\underline{s}) = (1,\underline{t})} c_{(2,\underline{s})}(X) E_{(2,\underline{s})}(X).
\eean
By~\eqref{eq:prop1} in Lemma~\ref{lem:poly_props}, the coefficient of $X^{\ell}$ is zero in $E_{(2,\underline{t})}(X)$ whenever $\ell \neq 0 \ (\textsl{mod} \ n_2)$. Hence for every $\underline{t} \in T^\prime_1$, there are of $\mu_2r_2$ monomials in $d_{(1,\underline{t})}(X)$. Evaluations of $d_{(1,\underline{t})}(X)$ at $n_1$ points in $A_{(1,\underline{t})}(X)$ give rise to an $[n_1,\mu_2r_2]$-codeword. Since the desired dimension of the middle code is $r_1$, we precode the message symbols such that the coefficients of $(r_2\mu_2 - r_1)$ highest degree monomials in $d_{(1,\underline{t})}(X)$ vanishes to zero. The polynomials $c_{(1,\underline{t})}(X)$ thus obtained corresponds to an $[n_1,r_1]$-middle code, and hence referred to as a {\em first level code polynomial}. We can write 
\bean
c_{(1,\underline{t})}(X) & = & \sum_{\underline{s}:\pi(2,\underline{s}) = (1,\underline{t})} P_1(c_{(2,\underline{s})}(X)) E_{(2,\underline{s})}(X), \ \ \underline{t} \in T^\prime_1 ,
\eean
where $P_1(\cdot)$ denotes the precoding transformation at the first level. In the next iteration, we compute $d_{(0,(1,0,0))} (X)$ and subsequently precode the message symbols by $P_0(\cdot)$ to reduce the dimension from $\bar{\mu}_1r_1$ to $k$ to obtain the {\em zeroth level code polynomial} $c_{(0,(1,0,0))} (X)$:
\bean
d_{(0,(1,0,0))} (X) \ = \ \sum_{\substack{\underline{s}:\pi(1,\underline{s}) = (0,(1,0,0)) \\ \underline{s} \in T^\prime_1}} c_{(1,\underline{s})}(X) E_{(1,\underline{s})}(X) \\
c_{(0,(1,0,0))} (X) \ = \ \sum_{\substack{\underline{s}:\pi(1,\underline{s}) = (0,(1,0,0)) \\ \underline{s} \in T^\prime_1}} P_0(c_{(1,\underline{s})}(X)) E_{(1,\underline{s})}(X),
\eean
The code polynomial $c(X)$ is identified as 
\beq \label{eq:codepoly} c(X) = c_{(0,(1,0,0))}(X). 
\eeq

\subsubsection{Evaluation of $c(X)$} The codeword $\underline{c} \ = \ (c(\theta) \mid \theta \in A)$ is obtained by evaluating the polynomial $c(X)$ at $n$ points taken from 
\beqn A  =  \bigcup_{\underline{t} \in T^\prime_1} A_{(1,\underline{t})}.
\eeqn
This completes the description of the construction. By the construction, it is clear that the dimension and the minimum distance of the code are given by 
\bean 
k & = & |\{ \ell \mid  \text{coefficient of } X^{\ell} \text{ in } c(X) \neq 0 \}| \\
d & \geq & n - \textsl{deg} (c(X)).
\eean

\begin{note} A principal construction in~\cite{TamBar} for codes with all-symbol locality, relies on a partitioning of the roots of unity contained in a finite field into a subgroup and its cosets.  The construction then identifies polynomials that are constant on each coset and makes use of these polynomials in the construction.   The approach adopted here is along similar lines.
\end{note}

\begin{example} In this example, we construct a code with $[n,k] = [24,14]$ having locality parameters $(n_1,r_1)=(12,8)$ and $(n_2,r_2) = (4,3)$, satisfying the divisibility condition. We can choose the finite field $\mathbb{F}_{p^m} = \mathbb{F}_{5^2}$. Let $\alpha$ be a primitive element of $\mathbb{F}_{5^2}$. We have $n_0 = n = 24$, $\mu_1 = \bar{\mu}_1 = 2$, and $\mu_2 = 3$. We set 
\bean
H_0 & = & \mathbb{F}^*_{5^2} \\
H_1 & = & \{1, \beta_1, \beta_1^2, \ldots, \beta_1^{11} \} \\
H_2 & = & \{1, \beta_2, \beta_2^2, \beta_2^{3},
\eean 
where $\beta_0 = \alpha$, $\beta_1 = \alpha^2$ and $\beta_2 = \alpha^6$. The relevant coset-tree can be computed as 
\bean 
A_{(0,(1,0,0))}  =   H_0, \ A_{(1,(1,1,0))} = H_1, \ A_{(1,(1,2,0))} = \beta_0 H_1\\
A_{(2,(1,t_1,t_2))} =  \beta^{t_1-1}_0 \beta^{t_2-1}_1 H_2 , \ 1 \leq t_1 \leq 2, 1 \leq t_2 \leq 3 .
\eean
Let us define the index sets $T_1 = \{(t_0,t_1,t_2) \mid t_0 = 1, 1 \leq t_1 \leq 2, t_2 = 0\}, T_2 = \{(t_0,t_1,t_2) \mid t_0 = 1, 1 \leq t_1 \leq 2, 1 \leq t_2 \leq 3 \}$. For every $\underline{t}=(1,t_1,0) \in T_1$, we set $s_1$ as the unique element in $\{1,2\}\setminus\{t_1\}$ and then we have
\bean
E_{(1,\underline{t})}(X) & = &  \left( \frac{X^{12} - (\beta^{s_1-1}_0)^{12} }{(\beta^{t_1-1}_0)^{12} - (\beta^{s_1-1}_0)^{12}} \right)  \\
& := & a_{\underline{t}} X^{12} + b_{\underline{t}} .
\eean
Similarly for every $\underline{t}=(1,t_1,t_2) \in T_2$, we set $\{s_1,s_2\} = \{1,2,3\}\setminus\{t_2\}$ and then we have
\bean
E_{(2,\underline{t})}(X) & = & \left( \frac{X^{4} - (\beta^{t_1-1}_0\beta^{s_1-1}_1)^{4} }{(\beta^{t_1-1}_0\beta^{t_2-1}_1)^{4} - (\beta^{t_1}_0\beta^{s_1-1}_1)^{4}} \right) \left( \frac{X^{4} - (\beta^{t_1-1}_0\beta^{s_2-1}_1)^{4} }{(\beta^{t_1-1}_0\beta^{t_2-1}_1)^{4} - (\beta^{t_1}_0\beta^{s_2-1}_1)^{4}} \right) \\
& := & e_{\underline{t}} X^{8} + f_{\underline{t}} X^{4} + g_{\underline{t}}.
\eean
There are $|T_2|=6$ message polynomials denoted by $\{ u_{\underline{t}}(X) = u_{\underline{t},0} + u_{\underline{t},1} X + u_{\underline{t},2} X^2  \mid \underline{t} \in T_2 \}$, each of degree $(r_2-1)=2$. The second level code polynomial for each $\underline{t} \in T_2$ corresponding to a $[4,3]$-local code is taken to be $c_{(2,\underline{t})} (X) \ = \ u_{\underline{t}}(X)$. In the next step, the first level code polynomial $\{c_{1,\underline{s}} (X)\}$ is constructed as 
\bean
c_{1,\underline{t}} (X) & = & \sum_{\underline{s}: s_1=t_1,s_2=1}^{s_2=2} P_1(c_{(2,\underline{s})}(X)) (e_{\underline{s}} X^{8} + f_{\underline{s}} X^{4} + g_{\underline{s}} ) .
\eean
for each of $\underline{t} \in T_1$. By virtue of the precoding $P_1(\cdot)$, the term $X^{10}$ vanishes and the resultant polynomial $c_{1,\underline{t}} (X)$ corresponds to a $[12,8]$-middle code. Subsequently, the zeroth level code polynomial is constructed as 
\bean
c_{0,(1,0,0)} (X) & = & \sum_{\underline{s}: s_1=1,s_2=0}^{s_1=2} P_0(c_{(1,\underline{s})}(X)) (a_{\underline{s}} X^{12} + b_{\underline{s}} ) .
\eean
Without precoding $P_0(\cdot)$, we would have obtained a polynomial of degree $21$ having $16$ monomials. Precoding wipes out the terms $\{ X^{21}, X^{20}\}$, and the resultant polynomial $c_{0,(1,0,0)}(X) =: c(X)$ of degree $18$ is the code polynomial consisting of $14$ monomials. Thus $k=14$, and $d \geq 6$. The codeword $\underline{c}$ is given by $\underline{c} = (c(\theta) \mid \theta \in H_0 )$.

It is of interest to look at the exponents of monomials in polynomials of each level. From each level, we pick a candidate polynomial $c(X)$, $c_1(X) := c_{(1,(1,1,0))}(X)$, $c_2(X) := c_{(2,(1,1,2))}(X)$.

\begin{figure}[h!]
	\begin{center}	
		\bean
			\begin{array}{|c|c|c|c|} \hline  
			\cancel{3} & 2 & 1 & 0 \\
			\hline
			\end{array} &
			\begin{array}{|c|c|c|c||c|c|c|c||c|c|c|c|} \hline 
			\cancel{11} & \cancel{10} & 9 & 8 & \cancel{7} & 6 & 5 & 4 & \cancel{3} & 2 & 1 & 0 \\
			\hline
			\end{array}
		\eean
	\end{center}
	\begin{center}	 
		\bean
			\begin{array}{|c|c|c|c|c|c|c|c|c|c|c|c|} \hline 
			\cancel{23} & \cancel{22} & \cancel{21} & \cancel{20} & \cancel{19} & 18 & 17 & 16 & \cancel{15} & 14 & 13 & 12 \\
			\hline
			\cancel{11} & \cancel{10} & 9 & 8 & \cancel{7} & 6 & 5 & 4 & \cancel{3} & 2 & 1 & 0 \\
			\hline
			\end{array}
		\eean
	\end{center}	
\caption{Illustration of the exponents of monomials in $c_{2}(X)$, $c_{1}(X)$ and $c(X)$ in order. The canceled exponents are those whose coefficients are fixed to zero by precoding. \label{fig:exponent}}
\end{figure}
\end{example}

The illustration in Fig.~\ref{fig:exponent} gives an equivalent simplistic description of the $[24,14,6]$ code. This works in general. Let $\textsl{Exp}(f)$ represent the ordered set of exponents of the monomials in a polynomial $f(X)$. By ordered set, we mean that the elements of the set are listed in the descending order. For example, $\textsl{Exp}(X^3+\alpha X+1) = \{3,1,0\}$. For an ordered finite set $S$ of non-negative integers and a positive integer $r \leq |S|$,  we define $\textsl{Trunc}(S, r)$ as the set comprising of the last $r$ elements of the set. Then we have that 
\bean
\textsl{Exp}(c_{2}) & = & \{r_2-1, r_2-2, \ldots, 0\} \ = \ \textsl{Trunc}(\mathbb{Z}_{n_2},r_2) \\
\textsl{Exp}(c_{1}) & = & \textsl{Trunc}( \bigcup_{j=0}^{\mu_2-1} ( jn_2+ \textsl{Exp}(c_{2}) ), r_1) \\
\textsl{Exp}(c) & = & \textsl{Trunc}( \bigcup_{j=0}^{\bar{\mu}_1-1} (jn_1+ \textsl{Exp}(c_{1})), k),
\eean
where $\mathbb{Z}_{n} = \{0,1,\ldots, n-1\}$. The set $\textsl{Exp}(c)$ is an equivalent simplistic description of the code. In terms of $\textsl{Exp}(c)$, we can write the parameters of the code as $k =  |\textsl{Exp}(c)|, \ \ d  \geq  n - \max (\textsl{Exp}(c))$.

\subsection{Locality Properties Of the Code}
In this section, we will show that the code satisfies locality constraints. Consider the case $c(y)$ is lost. We need to recover it accessing $r_1$ other symbols $\{c(y_1),c(y_2),\ldots, c(y_{r_1})\}$ that along with $c(y)$ are part of an $[n_1,r_1]$ punctured code. Without loss of generality, let us assume that $y \in A_{(1,(1,1,0))}$. Using \eqref{eq:prop3}, \eqref{eq:prop4} in Lemma~\ref{lem:poly_props}, we can write
\bean
c(X) E_{(1,(1,1,0))}(X) & = & P_0(c_{(1,(1,1,0))}(X)) E_{(1,(1,1,0))}(X).
\eean
Evaluations at $r_1$ out of $n_1$ points in $A_{(1,(1,1,0))}$ will help reconstruct $P_0(c_{(1,(1,1,0))}(X))$, since $\textsl{deg}(P_0(c_{(1,(1,1,0))}(X))) \leq (r_1-1)$. Then we can recover $c(y) = P_0(c_{(1,(1,1,0))}(y)) E_{(1,(1,1,0))}(y)$. The same argument can be used inductively to show that each symbol within an $[n_1,r_1]$-middle code can be recovered by $r_2$ out of some $n_2$ symbols. This establishes the existence of $[n_2,r_2]$-local codes.

\subsection{Optimality Of the Code \label{sec:optimality}}

\bthm \label{thm:optimal_1} The $[n,k,d]$-code with $[n_1,r_1,\delta_1]$-middle codes and $[n_2,r_2,\delta_2]$-local codes constructed in Sec.~\ref{sec:constrn} achieves the optimal minimum distance if $r_2 \mid r_1 \mid k$.
\ethm
\bpf Let $c(X)$ denote the code polynomial. The proof follows from counting $\mathbb{Z}_{n}$ in two different ways. We can write
\bea \label{eq:countzn}
n \ = \ |\mathbb{Z}_{n}| & = & |\textsl{Exp}(c)| + |\mathbb{Z}_{n} \setminus \textsl{Exp}(c)|.
\eea
We have that $|\textsl{Exp}(c)| = k$. On the other hand, since $r_2 \mid r_1 \mid k$, we can count the number of exponents that are truncated while forming $c$ as,
\bean
|\mathbb{Z}_{n} \setminus \textsl{Exp}(c)| & = & \left(\frac{k}{r_2} - 1\right)(\delta_2-1) \\
& &+ \left(\frac{k}{r_1} - 1\right)(\delta_1-\delta_2) + (d - 1)
\eean
Substituting back in \eqref{eq:countzn}, we conclude that the code is optimal.
\epf

\bthm \label{thm:optimal_2} The $[n,k,d]$-code with $[n_1,r_1,\delta_1]$-middle codes and $[n_2,r_2,\delta_2]$-local codes constructed in Sec.~\ref{sec:constrn} achieves the optimal minimum distance if the following conditions hold:
\ben
\item $d = n_2 + \delta_2$
\item $\frac{n}{n_1} = \left\lceil \frac{k}{r_1} \right\rceil, \ \frac{n}{n_2} = \left\lceil \frac{k}{r_2} \right\rceil + 1$
\een
\ethm
\bpf Let $c(X)$ denote the code polynomial. The proof is analogous to that of Thm.~\ref{thm:optimal_1}. The only difference lies in the count of $|\mathbb{Z}_{n} \setminus \textsl{Exp}(c)|$.  Since $d = n_2 + \delta_2$, we obtain that,
\bean
|\mathbb{Z}_{n} \setminus \textsl{Exp}(c)|  =  \left(\frac{n}{n_2} - 1\right)(\delta_2-1) + \left(\frac{n}{n_1} - 1\right)(\delta_1-\delta_2) \\
+ ((d - 1) - (\delta_2-1)) \\
=  \left(\frac{n}{n_2} - 2\right)(\delta_2-1) + \left(\frac{n}{n_1} - 1\right)(\delta_1-\delta_2) + (d - 1).
\eean
Hence the code is optimal if the second condition in the theorem holds. 
\epf
While Thm.~\ref{thm:optimal_1}, Thm.~\ref{thm:optimal_2} provide optimality conditions that can be generalized to hierarchical locality of arbitrary levels, a subject of discussion in Sec.~\ref{sec:chl-gen}, the next theorem characterizes the conditions for optimality for two-level hierarchy without imposing any restrictions.

\bthm \label{thm:optimal_3} The $[n,k,d]$-code with $[n_1,r_1,\delta_1]$-middle codes and $[n_2,r_2,\delta_2]$-local codes constructed in Sec.~\ref{sec:constrn} achieves the optimal minimum distance if 
\bea \label{eq:two-hier-cond}
\left\lceil \frac{k}{r_2} - \left(  \left\lceil \frac{k}{r_1} \right\rceil  - 1  \right)\frac{ r_1}{r_2} \right\rceil = \left\lceil \frac{k}{r_2} \right\rceil - \left(  \left\lceil \frac{k}{r_1} \right\rceil  - 1  \right) \left\lceil \frac{r_1}{r_2}\right\rceil .
\eea
\ethm
\bpf Let $c(X)$ denote the code polynomial. The proof is analogous to that of Thm.~\ref{thm:optimal_1}. It is possible to count the size of $\mathbb{Z}_{n} \setminus \textsl{Exp}(c)$ as

\bean
\lefteqn{\left( \left( \left\lceil \frac{r_1}{r_2} \right\rceil -1\right) (\delta_2 - 1) + (\delta_1 -1) \right) \left(  \left\lceil \frac{k}{r_1} \right\rceil  - 1  \right) } \\
& & + \left(  \left\lceil \frac{\left( k - \left(  \left\lceil \frac{k}{r_1} \right\rceil  - 1  \right) r_1 \right)}{r_2} \right\rceil - 1 \right) (\delta_2 -1 ) + (d-1).
\eean

The expression can be recast into the form
\bean
\lefteqn{\left(  \left\lceil \frac{k}{r_1} \right\rceil  - 1  \right) ( \delta_1 - \delta_2) + (d-1) + } \\
&&\left(  \left\lceil \frac{k}{r_2} - \left(  \left\lceil \frac{k}{r_1} \right\rceil  - 1  \right)\frac{ r_1}{r_2} \right\rceil  +  \left(  \left\lceil \frac{k}{r_1} \right\rceil  - 1  \right) \left\lceil \frac{r_1}{r_2}\right\rceil - 1 \right) (\delta_2 -1),
\eean
thus leading to a value of $d$ as in
\bean
d & = & n - k + 1 - \left(  \left\lceil \frac{k}{r_1} \right\rceil  - 1  \right) ( \delta_1 - \delta_2) \\ 
& - & \left(  \left\lceil \frac{k}{r_2} - \left(  \left\lceil \frac{k}{r_1} \right\rceil  - 1  \right)\frac{ r_1}{r_2} \right\rceil + \left(  \left\lceil \frac{k}{r_1} \right\rceil  - 1  \right) \left\lceil \frac{r_1}{r_2}\right\rceil - 1 \right) (\delta_2 -1) .
\eean
Comparing against the upper bound in~\eqref{eq:bound}, we conclude that the code is optimal if~\eqref{eq:two-hier-cond} holds.
\epf

\section{Generalization To $h$-Level Hierarchy \label{sec:chl-gen}}

In Sec.~\ref{sec:chl-2}, we considered codes with hierarchical locality where the hierarchy had two levels. Here we extend the notion to $h$-level hierarchy where $h$ is an arbitrary number. 
\bdefn An $[n,k,d]$ linear code $\mathcal{C}$ is a {\em code with $h$-level hierarchical locality} having locality parameters $[(r_1,\delta_1),(r_2,\delta_2), \ldots (r_h,\delta_h)]$ if for every symbol $c_i, 1 \leq i \leq n$, there exists a punctured code $C_i$ such that $c_i \in \textsl{Supp}(C_i)$ and the following conditions hold:
\ben
\item $\textsl{dim}(C_i) \leq r_1$ ,
\item $d_{\text{min}}(C_i) \geq \delta_1$ ,
\item $C_i$ is a code with $(h-1)$-level hierarchical locality having locality parameters $[(r_2,\delta_2), (r_3,\delta_3), \ldots (r_h,\delta_h)]$.
\een
Code with $1$-level hierarchical locality is defined to be code with locality.
\edefn

The punctured code $C_i$ associated with $c_i$ is referred to as {\em local code of level-$1$}. In fact, each symbol is associated with a bunch of local codes, each of level-$i$, $i=1,2,\ldots, h$. In the previous section, we studied codes with $2$-level hierarchical locality. 

\subsection{An Upper Bound on The Minimum Distance}

\bthm \label{thm:hbound} Let $\mathcal{C}$ be an $[n,k,d]$-linear code with $h$-level hierarchical locality having locality parameters $[(r_1,\delta_1),(r_2,\delta_2), \ldots, (r_h,\delta_h)]$. Then
\bea \label{eq:bound_general}
\nonumber d & \leq & n-k+1 - \sum_{\ell = 1}^{\ell = h-1}\left(\left( \left\lceil \frac{k}{r_\ell} \right\rceil - 1 \right) (\delta_\ell-\delta_{\ell+1})\right) \\
&  & - \left( \left\lceil \frac{k}{r_h} \right\rceil - 1 \right)(\delta_h -1).
\eea
\ethm
\bpf The proof is a straightforward extension of that of Thm.~\ref{thm:bound}. The algorithm~\ref{alg:hbound}, (see flow chart in Fig. \ref{fig:flowchart_extension_proper}) identifies a $(k-1)$-dimensional punctured code $\mathcal{C}_s$ of $\mathcal{C}$, having large support. Then we will use the Fact in~\eqref{eq:fact}. 
The algorithm identifies a level-$1$ code that accumulates rank, and subsequently visits a level-$2$ code from within that, and continues recursively upto reaching a level-$(h-1)$ code that accumulates rank. Then it picks up all the level-$h$ codes that accumulate rank. If no more level-$h$ codes can be picked up, it steps back one level up, and finds a new level-$(h-1)$ code that accumulates rank. This can be viewed as a depth-first search for rank-accumulating level-$h$ codes. At each level, incremental support is added to the variable $\Psi$. Vaguely speaking, the incremental support that is added at each level depends on the minimum distance of the code at that level.
\begin{algorithm}
	\caption{For the proof of Thm.~\ref{thm:hbound} \label{alg:hbound}}
	\begin{algorithmic}[1]
		\State Let $i_1=0, i_2=0, \ldots, i_h =0;  \Psi = \Phi, \ell = 1; $
		\State $M_{0}= \mathcal{C}, M_{1}= \Phi, M_{2}= \Phi, \ldots,  M_{h}= \Phi$
		\While {($ \textsl{rank}(G|_{\Psi}) < k$)}
		\If {($\exists$ a level-$\ell$ code $L_{i_\ell} \in M_{\ell - 1} $ such that $\textsl{rank}(G|_{\Psi \cup \textsl{Supp}(L_{i_\ell})}) > \textsl{rank}(G|_{\Psi})$)}
		\State $M_{\ell} = L_{i_\ell}$
		\If {($\ell \text{ equals } h $)}
		\State $\Psi = \Psi \cup \textsl{Supp}(M_{h})$
		\State $S_{\text{last}} = \textsl{Supp}(M_{h})$
		\State $i_{h} = i_{h} + 1$
		\Else					
		\State $\ell = \ell + 1$
		\EndIf					
		\Else 
		\State $\ell = \ell - 1$;
		\State $\Psi = (\Psi \setminus S_{\text{last}}) \cup T_{\ell}$
		\State $S_{\text{last}} = T_{\ell}$
		\State $i_{\ell} = i_{\ell} + 1$
		\EndIf					
		\EndWhile
	\end{algorithmic}
\end{algorithm} 
Let $a_{i_h}$ denote the incremental rank and $s_{i_h}$ denote the incremental support while adding a level-$h$ code $L_{i_h}$. By the algorithm, $a_{i_h} > 0$. The set $S_{i_h}$ denotes the support of the level-$h$ code $L_{i_h}$. The set $T_{\ell}$ denotes the incremental support of the level-$\ell$ code $M_{\ell}$, along with the columns from the last code that accumulated rank. It will contain at least $(d_\ell - 1)$ columns in addition to the incremental rank. Let $i_{\ell_\text{end}}$ denote the final value of the variables $i_{\ell}, 1 \leq \ell \leq h$  before the algorithm terminates. Since $a_{i_h} > 0$, we can get non-trivial lower bounds on $s_{i_h}$ and $t_{\ell} := |T_{\ell}|$ quite similar to the proof of Thm.~\ref{thm:bound}. The rank is accumulated to $k$ after adding the last local code $L_{i_{h_{\text{end}}}}$. By this time, we would have also visited $i_{\ell_{\text{end}}}$ level-$\ell$ codes. Hence clearly,
\bean
i_{\ell_{\text{end}}}  \geq  \left\lceil \frac{k}{r_\ell} \right\rceil,  \ \ 1 \leq \ell \leq h    
\eean
After adding $i_{h_{\text{end}}}-1$ level-$h$ codes, we would have accumulated rank that is less than or equal to $(k-1)$. Hence we can always pick $(k-1) - \sum_{i_h=1}^{i_{h_{\text{end}}}-1} a_{i_h}$ columns from $L_{i_{h_{\text{end}}}}$ so that the total rank accumulated becomes $(k-1)$. The resultant punctured code is identified as $\mathcal{C}_s$. Following a similar line of arguments as in the proof of Thm.~\ref{thm:bound}, we can get an estimate on $\textsl{Supp}(\mathcal{C}_s)$ as
\bean 
|S| & \geq & (k-1) +   \sum_{\ell = 1}^{\ell = h-1}\left(\left( \left\lceil \frac{k}{r_\ell} \right\rceil - 1 \right) (d_\ell-d_{\ell+1})\right) \\ 
&  & + \left( \left\lceil \frac{k}{r_h} \right\rceil - 1 \right)(d_h -1).
\eean
Hence the theorem follows. 
\epf

\subsection{Code Construction For All-Symbol Locality}
The construction in Sec.~\ref{sec:constrn} of the main text can be generalized to construct codes with $h$-level hierarchical locality containing $[n_i,r_i, \delta_i]$ codes as $i$-th level code for each $i=1,2,\ldots,h$. Here also, we require to satisfy a divisibility condition $n_h \mid n_{h-1} \mid \cdots \mid n$. The generalization is straightforward, and it boils down to finding a finite field $\mathbb{F}_{p^m}$ such that $n_h \mid n_{h-1} \mid \cdots \mid n_1 \mid (p^m-1)$, and $n \leq (p^m-1)$. Then we can find a subgroup chain $H_h \subset H_{h-1} \subset \cdots \subset H_0 = \mathbb{F}^*_{p^m}$. This allows us to create a coset-tree of depth $h$, and the code construction follows naturally. It can also be proved that the construction thus obtained will be optimal in terms of minimum distance if either of the two conditions holds:
\ben
\item $r_h \mid r_2 \mid \cdots \mid r_1 \mid k$.
\item $d = n_h + \delta_h , \frac{n}{n_h} = \left\lceil \frac{k}{r_h} \right\rceil + 1, \frac{n}{n_i} = \left\lceil \frac{k}{r_i} \right\rceil, \ \forall i = 1,2, \ldots, h-1$. 
\een

\bibliographystyle{IEEEtran}
\bibliography{isit2015_arxiv}

\appendices

\section{Flowcharts For Algorithms Used to Derive Bounds\label{app:flowcharts}}

\begin{figure*}[ht]
  \centering
  \subfigure[Algorithm \ref{alg:bound}]{\label{fig:flowchart_proper}\includegraphics[width=2.6in]{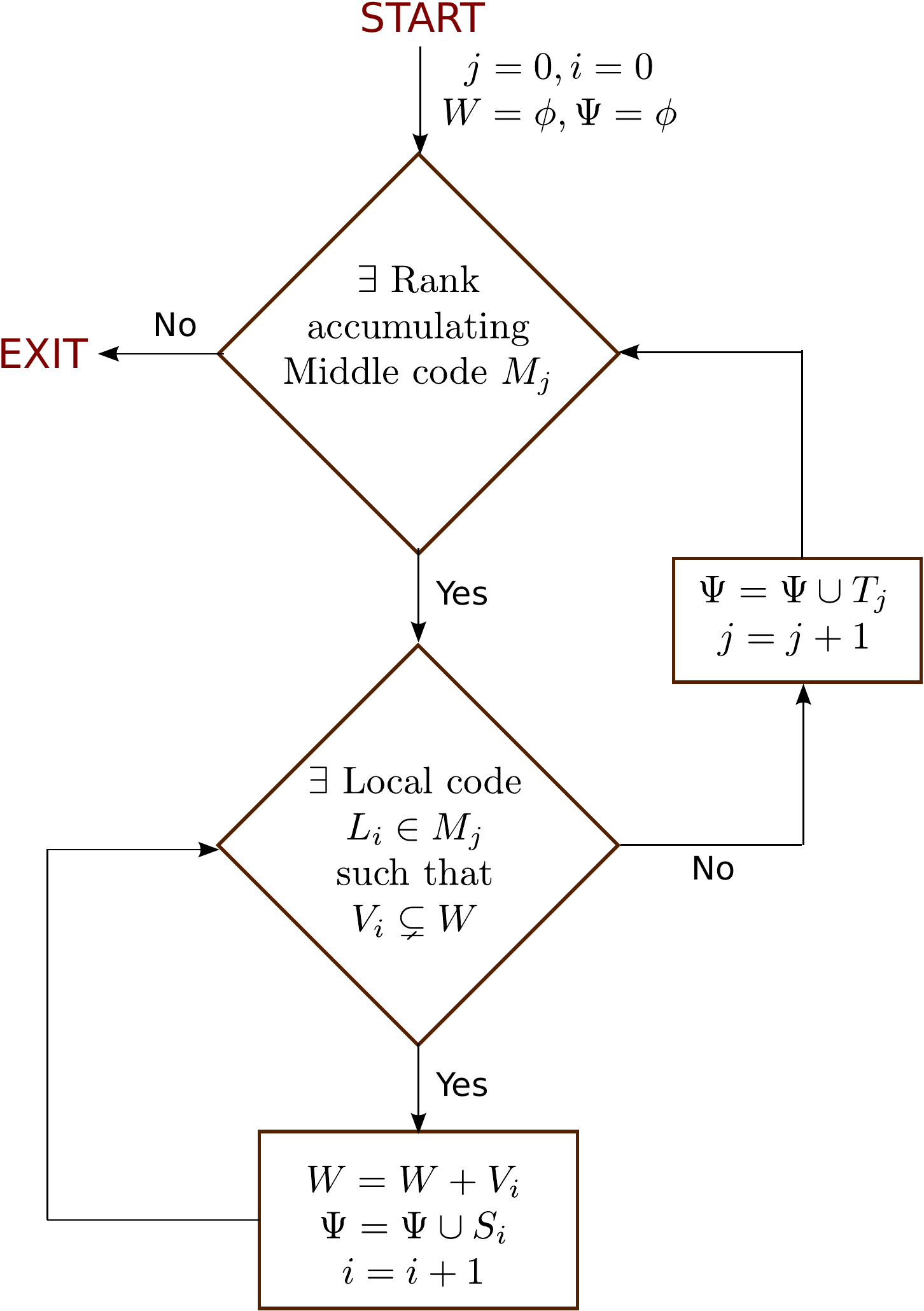}}
  \hspace{0.2in}
  \subfigure[Algorithm \ref{alg:hbound}]{\label{fig:flowchart_extension_proper}\includegraphics[width=2in]{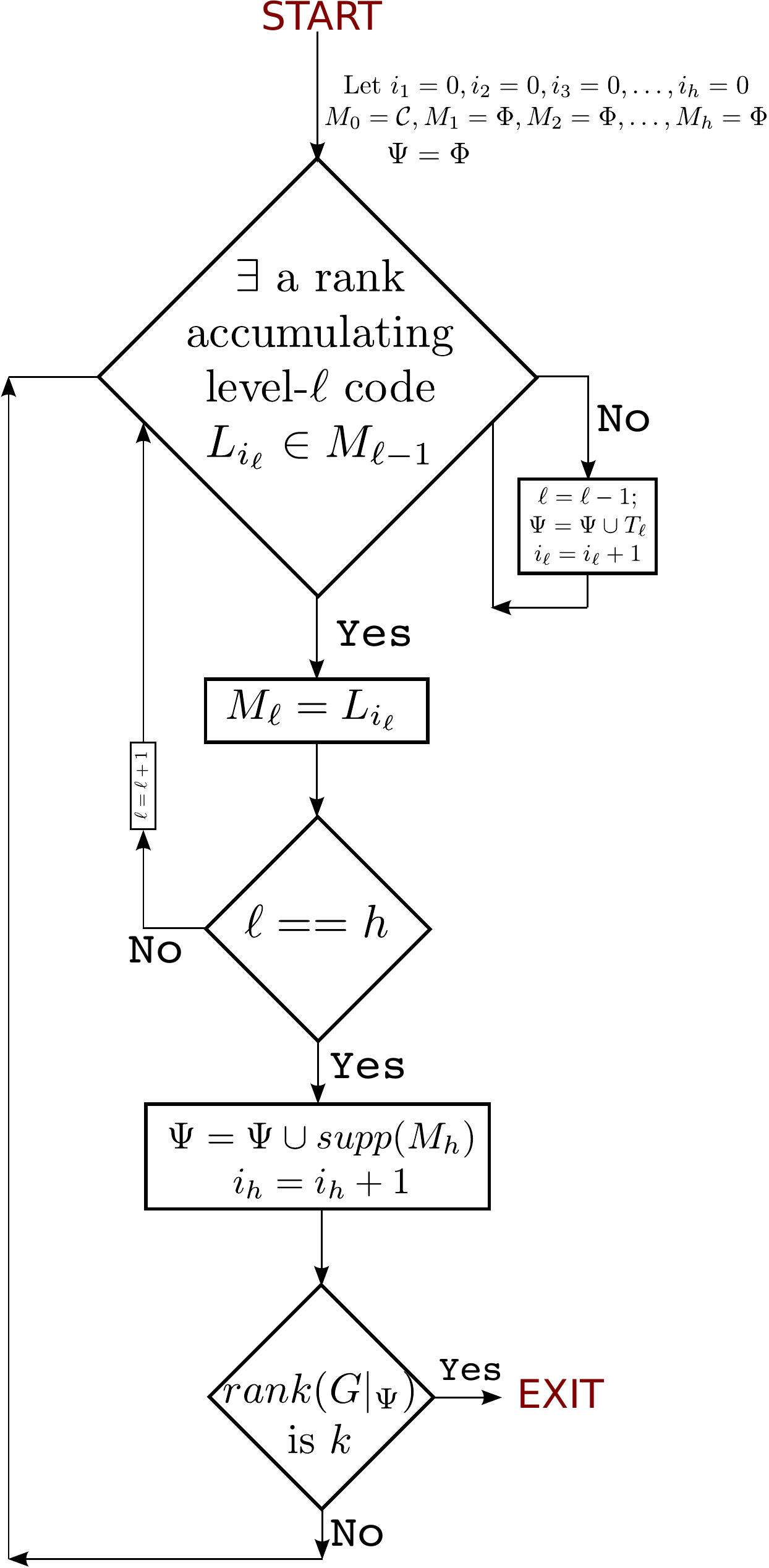}}
  \caption{Flowchart for algorithms uses for proving the bounds on minimum distance.}
  \label{fig:flowcharts}
\end{figure*}

The flow charts for algorithms \ref{alg:bound} and \ref{alg:hbound} are shown in Fig. \ref{fig:flowchart_proper} and Fig. \ref{fig:flowchart_extension_proper} respectively.
%

\section{Existence Of Required Field\label{app:num_theory_prime}}

First we will show that there exists a prime $p$ such that $n_1 \mid p - 1$. By Dirichlet’s theorem, if $a$ and $d$ are two co-prime numbers, then the sequence $a, a + d, a + 2d, \ldots$ will contain infinitely many primes. By setting $a = n_1+1$ and $d=n_1$, we observe that there are infinitely many primes of the form $(n_1+1) + \ell n_1$, i.e. of the form $(\ell + 1)n_1 + 1$. Thus we obtain a prime $p$ such that $n_1 \mid p - 1$. If $n \leq p - 1$, we are done. If not, pick a sufficiently large $m$ such that $n < p^m$. Since $n_1 \mid p -1$, we must also have $n_1 \mid p^m - 1$. 

\section{Proof Of Lemma \ref{lem:poly} }
It is sufficient to verify that 
\bea
\label{eq:cond1} q_{(i,\underline{t})}(X) & \mid & E_{(i,\underline{t})}(X) \\
\label{eq:cond2}  \sum_{\substack{t_i = 1 \\ t_j \text{fixed for} j \neq i}}^{\mu_i} E_{(i,\underline{t})}(X) & = & 1 \ \text{mod} \ (X^{n_{i+1}} -  \gamma_{(i-1,\underline{\tau})}^{n_{i-1}}) .
\eea
where $\underline{\tau}$ is the unique element such that $(i-1,\underline{\tau}) = \pi(i,\underline{t})$ for every $\underline{t}$ participating in the summation. For every such $\underline{t}$, the roots of $q_{(i,\underline{t})}(X)$ are precisely 
\beq \label{eq:parent}
\Lambda_{(i,\underline{t})} = \gamma_{(i-1,\underline{\tau})} H_{i-1} \setminus A_{(i,\underline{t})}.
\eeq
It can also be checked that $E_{(i,\underline{t})}(X)$ evaluates to zero at any point in $\Lambda_{(i,\underline{t})}$. Hence $q_{(i,\underline{t})}(X)) \mid E_{(i,\underline{t})}(X)$. It can also be seen that at any point $y \in \gamma_{(i-1,\underline{\tau})} H_{i-1}$, all except one term in the L.H.S. of \eqref{eq:cond2} evaluates to zero, and the remaining term evaluates to $1$. Hence \eqref{eq:cond2} holds, thereby completing the proof.

\end{document}